\documentclass[aps,prl,reprint,superscriptaddress,showpacs]{revtex4-1} 

\usepackage{amsmath,amsthm,amsfonts,amssymb,bm}
\usepackage{graphicx,array,MnSymbol}
\usepackage[usenames,dvipsnames]{xcolor}
\definecolor{OceanBlue}{rgb}{0,0.35,0.7} 
\usepackage{hyperref}
\hypersetup{colorlinks=true,allcolors=OceanBlue}

\theoremstyle{definition}

\theoremstyle{remark}

\newcommand{\lp}{\left(}  
\newcommand{\rp}{\right)} 
\newcommand{\lb}{\left[}   
\newcommand{\rb}{\right]}  

\newcommand{\mr}{\mathrm} 
\newcommand{\mb}{\mathbf} 
\newcommand {\cl}{\mathcal} 
\newcommand {\tsf} [1]{\textsf{#1}} 
\newcommand{\vectsym}{\boldsymbol}  
\DeclareMathOperator{\Tr}{Tr\,}            

\newcommand*\diff{\mathop{}\!\mathrm{d}}
\DeclareMathOperator{\trans}{\textsf{T}} 

\newcommand{\brho}{\vectsym{\rho}} 
\newcommand{\bsigma}{\vectsym{\sigma}} 
\newcommand{\bd}{\mb{d}} 

\newcommand {\ket}[1] {\left|{#1}\right\rangle}

\newcommand {\bra}[1] {\langle{#1}|}

\newcommand{\abs}[1]{\left | #1 \right |}   
\newcommand{\mean}[1]{\left\langle #1 \right\rangle}  

\begin{document}
\title{Far-field Superresolution of Thermal Electromagnetic  Sources  at the Quantum Limit}
\author{Ranjith Nair}
\email[Corresponding author: ]{elernair@nus.edu.sg}
\affiliation{Department of Electrical and Computer Engineering, National University of Singapore, 4 Engineering Drive 3, Singapore 117583}
\author{Mankei Tsang}
\affiliation{Department of Electrical and Computer Engineering, National University of Singapore, 4 Engineering Drive 3, Singapore 117583}
\affiliation{Department of Physics, National University of Singapore, 2 Science Drive 3,
Singapore 117551}

\date{\today}
\begin{abstract} 
We obtain the ultimate quantum limit for estimating the transverse separation of two thermal point sources using a given imaging system with limited spatial bandwidth. We show via the quantum Cram\'er-Rao bound that,  contrary to the Rayleigh limit in conventional direct imaging, quantum mechanics does not mandate any loss of precision in estimating even deep sub-Rayleigh separations. We propose two coherent measurement techniques, easily implementable using current linear-optics technology, that approach the quantum limit over an arbitrarily large range of separations. Our bound is  valid for arbitrary source strengths, all regions of the electromagnetic spectrum, and for any imaging system with an inversion-symmetric point-spread function. The measurement schemes can be applied  to microscopy, optical sensing, and astrometry at all wavelengths.
\end{abstract}
\maketitle

The Rayleigh criterion for resolving two incoherent optical point sources \cite{Ray1879xxxi,*BW99Principles}  is the most widely used benchmark for the resolving power of an imaging system. According to it, the sources can be resolved by direct imaging only if they are separated by at least the diffraction-limited spot size of the point-spread function of the imaging system. While the criterion is heuristic and does not take into account the intensity of the sources or the measurement shot noise, recent work \cite{BDD+99,VDD+02,RWO06,CWO16} has made it rigorous by taking as resolution measure   the classical Cram\'er-Rao lower bound (CRB) of estimation theory \cite{Cra16,*Rao45,*VanTreesI} on the mean squared error (MSE) of any unbiased estimate of the separation of the sources using spatially-resolved image-plane photon counting. These works showed that if the detected average photon number per mode $N_s \ll 1$, the MSE of any unbiased estimator based on direct imaging  diverges as the source separation decreases to zero over an interval comparable to the Rayleigh limit. This phenomenon, dubbed {Rayleigh's curse} in  \cite{TNL16}, stems from the indistinguishability between the photons coming from the two sources and imposes a fundamental limitation of direct imaging in resolving sources much closer than the spot size, even when the measured photon number is taken into account. Recent developments in far-field microscopy \cite{WS15} sidestep Rayleigh's curse by preventing multiple sources from emitting simultaneously, but control over the emission properties of sources is unavailable in target sensing or astronomical imaging.

While the development of novel quantum states of light and measurement techniques has given rise to the vast field of quantum imaging \cite{LGB02,*Shi07,*Kol07quantum}, fundamental quantum limits in resolving two \emph{incoherent} sources have been largely neglected since the early days of quantum estimation theory \cite{Hel73,Hel76}. Recently, the coherent \cite{Tsa15} and incoherent \cite{TNL16} two-source resolution problems were revisited using the quantum Cram\'er-Rao bound (QCRB) \cite{Hel76,Hol11} that accounts for all (unbiased) measurement techniques allowed by quantum mechanics. Under a weak-source assumption similar to that in \cite{BDD+99,VDD+02,RWO06,CWO16}, it was found in \cite{TNL16} that the QCRB showed no dependence on the separation of the sources. Linear optics-based measurements   that approach the bound were also proposed \cite{TNL16,NT16}.  Subsequent  demonstrations of superresolution \cite{TDL16,YTM+16,TFS16,PSH+16} have substantiated the feasibility of these proposals. Nevertheless, the classical treatments \cite{BDD+99,VDD+02,RWO06,CWO16} and the quantum treatment \cite{TNL16} neglect multi-photon coincidences and bunching, phenomena that figure prominently in quantum optics \cite{MW95}. While such an approximation leads to correct conclusions for weak sources, e.g., at optical frequencies \cite{Zmu03b,*Tsa11}, it is problematic for intense sources, e.g., in the microwave to far-infrared regimes, for high-temperature astronomical sources, and for optical demonstrations using pseudothermal light generated from laser sources \cite{ENT71}. As such, a quantum-optically rigorous derivation of the resolution limit is as yet unavailable.

In this paper, we solve these problems and obtain the QCRB for estimating the separation of two thermal point sources of arbitrary strength using  rigorous quantum optics and estimation theory, and show that  resolution is not fundamentally compromised at sub-Rayleigh separations.  We then propose two schemes that approach the QCRB. The finite spatial-mode demultiplexing
(fin-SPADE) scheme performs photon counting in a finite number of suitably chosen transverse spatial modes of the field. The interferometric pixelated superlocalization by image inversion interferometry (pix-SLIVER) scheme uses pixelated detector arrays in the two interferometer outputs. The two schemes approach the QCRB over greater ranges of the separation as the number of accessed modes (fin-SPADE) or the number of pixels (pix-SLIVER) is increased.

\begin{figure}[htbp]
\centering\includegraphics[trim=0mm 29mm 0mm 44mm,clip=true,width=0.8\columnwidth]{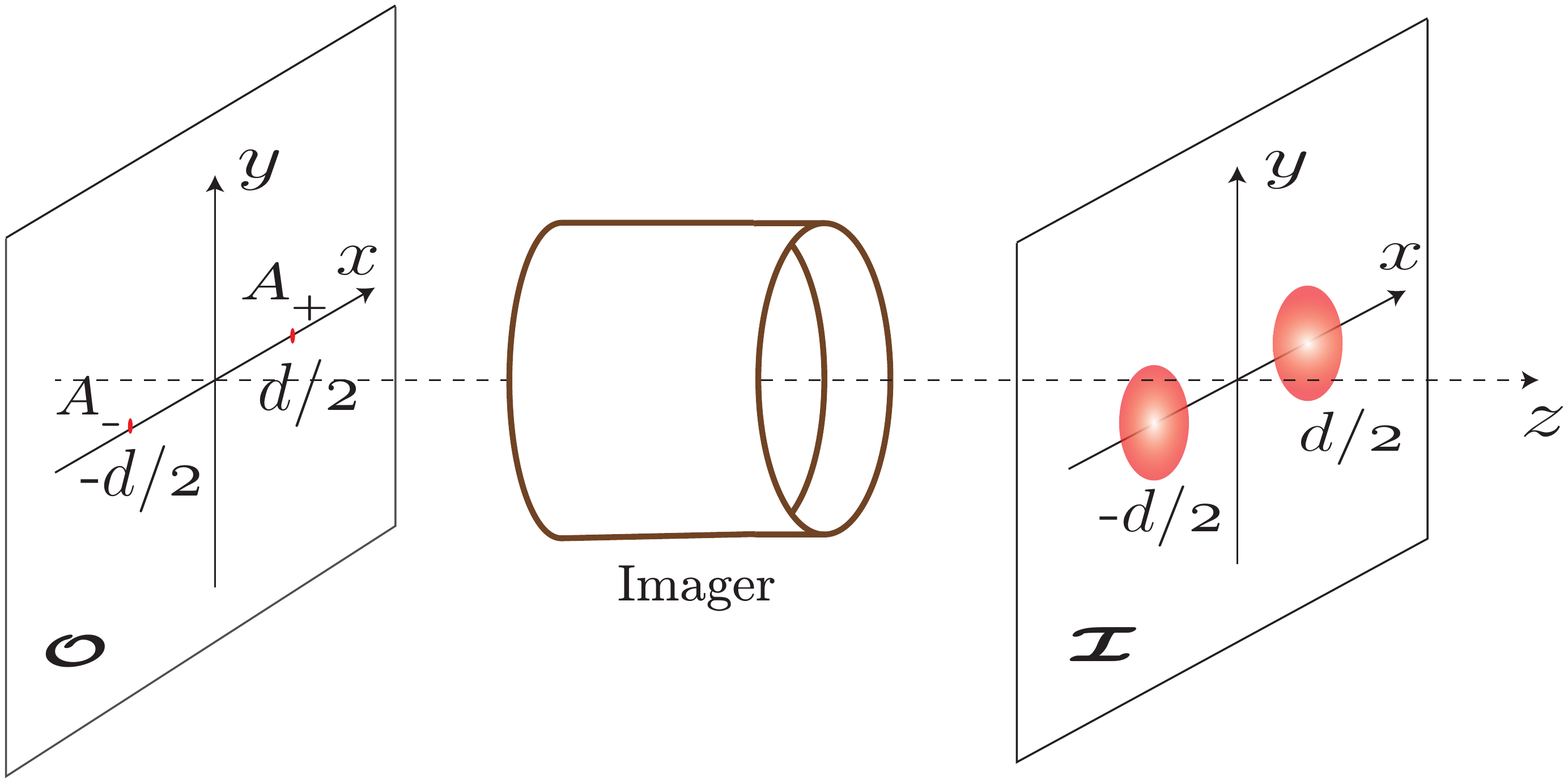}
\caption{\label{fig:system} A spatially-invariant imaging system: Point sources at $(\pm d/2,0)$ of the object plane $\cl{O}$ have images centered at $(\pm d/2,0)$ of the image plane $\cl{I}$ but spread out by the PSF of the system.}
\end{figure}

\paragraph*{Source and system model:} Consider two thermal point sources being imaged under paraxial conditions by a spatially-invariant  unit-magnification imaging system (Fig.~\ref{fig:system}) -- such an assumption entails no essential loss of generality  \cite*{[{}][{, Sec.~5.3.}]Goo05Fourier}.  We assume that the system's amplitude point-spread function (PSF)  $\psi(\brho)\, \lp \int_{\cl I}\diff\brho \abs{\psi(\brho)}^2 =1\rp $ is inversion-symmetric, i.e., $\psi(-\brho) = \psi(\brho)$, where $\brho = (x,y)$ is the transverse coordinate in the image plane $\cl{I}$. Most imaging systems, e.g., those with circular or rectangular entrance pupils, satisfy this assumption \cite{Goo05Fourier}.

Two incoherent thermal point sources, each of effective strength (average photon number) $N_s$ \footnote{To be precise, $N_s$ is the average number of photons from each source \emph{reaching the image plane}, allowing us to write Eqs.~\eqref{eq:coherent}-\eqref{eq:eigenfunc}.}, are described by a pair of dimensionless amplitudes $A = (A_+,A_-) \in \mathbb{C}^2$  with the probability density \cite{Goo85Statistical,MW95}:
\begin{align}
P_{N_s}(A) =  \lp{\pi N_s}\rp^{-2} \, \exp\lb- \lp\abs{A_+}^2 + \abs{A_-}^2\rp /{N_s}\rb.
\end{align}
 In order to focus on the essential physics of the problem,  we assume that the centroid (midpoint) of the sources is imaged at  the optical axis and that the line joining the sources is aligned with the $x$-axis, so that images of the sources are centered at $\bm{d}_{\pm}=(\pm d/2, 0)$ respectively  in the image plane. Estimating the centroid of two incoherent sources by direct imaging is subject to much less stringent bounds than the separation \cite{Hel76,BDD+99,TNL16} and may be done using a portion of the available signal \cite{TNL16}.  We also assume that  a single quasimonochromatic temporal mode $\xi(t) \lp\int_0^T \diff t \abs{\xi(t)}^2 = 1\rp$ is excited over the observation interval $[0,T]$. Extensions to multiple temporal modes can be made using standard techniques \cite{Hel76}.

\begin{figure}[htbp]
\centering\includegraphics[trim=25mm 84mm 24mm 86mm,clip=true,width=0.8\columnwidth]{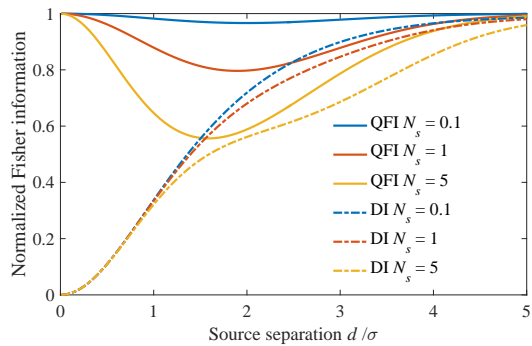}
\caption{\label{fig:QFI} (Color online) The QFI of Eq.~\eqref{eq:QFI} (solid lines), the lower bound of Eq.~\eqref{eq:FIlb} (dash-dotted lines) on spatially- and number-resolved direct imaging (DI) for the Gaussian PSF \eqref{eq:gaussian}. The plots are normalized to the respective maximum values $N_s/2\sigma^2$ of the QFI and are independent of the PSF half-width $\sigma$. }
\end{figure}
Conditioned on the value of $A$, the electromagnetic field in the image plane, described by the positive-frequency field operator $\hat{E}(\brho,t)$ \footnote{We assume a single polarization and that the quasimonochromatic scalar field operator  $\hat{E}(\brho,t)$ has been cast in units of $\sqrt{\mathrm{photons}\cdot\mathrm{m}^{-2}\cdot\mathrm{s}^{-1}}$},  is in a pure coherent state $\ket{\psi_{A,d}}$ that is an eigenstate of $\hat{E}(\brho,t)$  with the eigenfunction $\psi_{A,d}(\brho,t)$ given by:
\begin{align}
&\hat{E}(\brho,t) \, \ket{\psi_{A,d}} = \psi_{A,d}(\brho,t)\,\ket{\psi_{A,d}};  \label{eq:coherent}\\
&\psi_{A,d}(\brho,t)= \lb A_+\,\psi(\brho - \bm{d}_+) + {A}_- \, \psi(\brho - \bm{d}_-) \rb\, \xi(t), \label{eq:eigenfunc}
\end{align}
where we have used the spatial invariance of the imaging system to write \eqref{eq:eigenfunc}. The unconditional quantum state $\rho_d$ then has the $P$-representation \cite{MW95}:
\begin{align}\label{eq:state}
\rho_d =  \int_{\mathbb{C}^2} \diff^2 A_+ \diff^2 A_-  \, P_{N_s}(A) \ket{\psi_{A,d}}\, \bra{\psi_{A,d}}.
\end{align}

\paragraph*{Fundamental quantum bound:} The quantum Fisher information (QFI) $\cl{K}_d$ of the state family $\{\rho_d\}$ determines the quantum Cram\'er-Rao bound (QCRB) 
\begin{align} \label{eq:qcrbdef}
\mathbb{E}\lb \check{d} - d\rb^2 \geq \cl{K}_d^{-1}
\end{align}
on the MSE of \emph{any} estimator $\check{d}$ of the separation derived from an unbiased measurement POVM \cite{Hel76,Hol11,Hay06}. Our derivation of  $\cl{K}_d$  proceeds by calculating the quantum fidelity $F(\rho_{d_1},\rho_{d_2}) = \Tr\sqrt{\sqrt{\rho_{d_1}} \rho_{d_2} \sqrt{\rho_{d_1}}}$ between the (noncommuting) states \eqref{eq:state} for two neighboring separations $d_1$ and $d_2$ and employing the relation
\begin{align} \label{eq:filqcrb}
\cl{K}_d = 8\times \lim_{d_1,d_2 \rightarrow d} \frac{1-F(\rho_{d_1},\rho_{d_2})}{(d_1 -d_2)^2}
\end{align}
between the fidelity and the QFI \cite{BC94,Hay06}. The details of the derivation are given in the Appendix, with the result:
\begin{align} \label{eq:QFI}
\cl{K}_d  = - 2\beta(0)N_s - 2\gamma^2(d) \lb\frac{(1+ N_s)N_s^2}{(1+N_s)^2 - \delta^2(d) N_s^2} \rb,
\end{align}
where
\begin{align}
\delta(d) = \int_{\cl I}\diff\brho\,\psi^*(\brho)\psi(\brho-(d,0))
\end{align}
is the overlap function of the PSF for translations in the $x-$direction, $\gamma(d) = \partial \delta(d)/\partial d$, and $\beta(d) =  \partial \gamma(d)/\partial d$ \footnote{Inversion symmetry of the PSF entails that $\delta(d) =\delta(-d) = \delta^*(d)$ \cite{Note3}. Note that $\delta(d) \leq \delta(0) = 1$, so that $\gamma(0)=0$}.  In particular, 
\begin{align}
-\beta(0) =  \int_{\cl I} \diff\brho\, \abs{\frac{\partial \psi(\brho)}{\partial x}}^2 \equiv (\Delta k_x^2),
\end{align}
the mean-squared spatial bandwidth of the PSF in the $x$-direction, and is independent of orientation for circular-symmetric PSFs. 
\begin{figure}[tbp]
\centering\includegraphics[trim=10mm 17.5mm 10mm 4mm,width=0.7\columnwidth,clip=true]{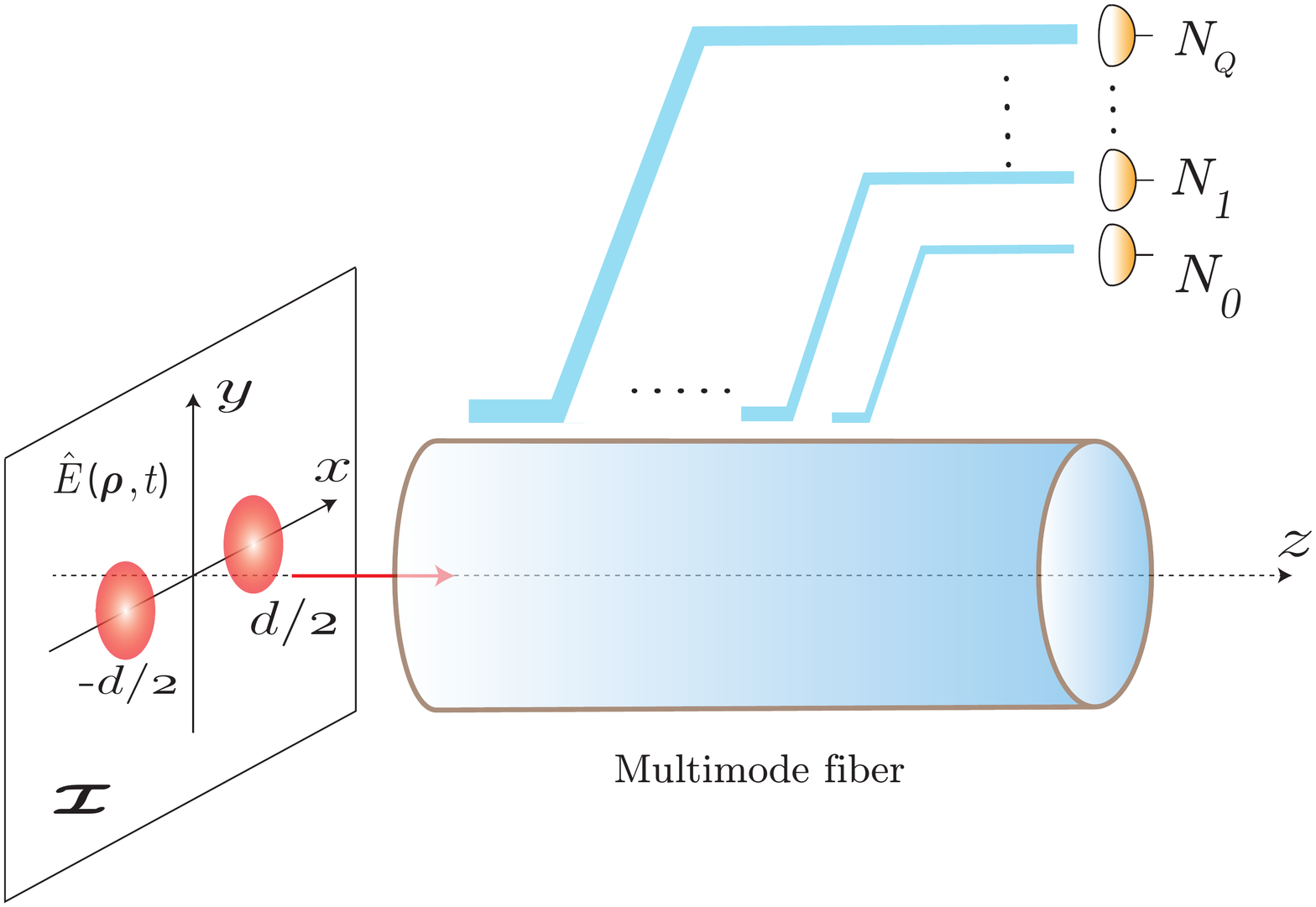}
\caption{\label{fig:SPADE}Fin-SPADE: The image-plane field is coupled into a multimode fiber and separated into its components in the Hermite-Gaussian TEM${}_{q0}$  modes of order $0 \leq q \leq Q$ by evanescent coupling to single-mode fibers supporting those modes. Detectors record the photon number in each mode. } 
\end{figure}
The first term in \eqref{eq:QFI} -- identical to the result in \cite{TNL16} -- is independent of $d$ and dominates in the $N_s \ll 1$ regime. For arbitrary $N_s$, this value is attained in the large-$d$ limit ($\gamma(d) \rightarrow 0$ as $d \rightarrow \infty$) but also for $d=0$, so that Rayleigh's curse is evaded.  The QFI suffers a dip at intermediate values whose relative depth increases with increasing $N_s$.  This is the net effect of  correcting the overestimation of the single-photon probability and neglect of multi-photon events in the weak-source model of \cite{TNL16}.  The QFI \eqref{eq:QFI} and a lower bound on the FI of spatially-resolved direct detection (see following)  are shown in Fig.~\ref{fig:QFI} for a system with the circular Gaussian PSF
\begin{align} \label{eq:gaussian}
\psi_G(\brho) ={(2\pi\sigma^2)^{-1/2}}\exp\lb-{\abs{\brho}^2}/\lp{4\sigma^2}\rp\rb,
\end{align}
for which $-\beta(0) =  {1}/{(4\sigma^2)}$.  

 Theoretical results guarantee the existence of multi-step POVMs that attain the QFI \cite{Hay05asymptotic,*Fuj06}, but we now give two linear-optics schemes that  closely approach it.

\paragraph*{Fin-SPADE:} For a system with the Gaussian PSF \eqref{eq:gaussian}, consider the separation of the image-plane field $\hat{E}(\brho,t)$ into its components in the TEM$_{q0}$ Hermite-Gaussian (HG) basis \cite{Yar89QuantumElect} $\{\psi_{q0}(\brho)\}_q$ with $\psi_G(\brho)\equiv \psi_{00}(\brho)$, followed by number-resolved but not necessarily time-resolved photon counting over $[0,T]$ in each of the modes with order $0 \leq q \leq Q$. The coupling to the TEM$_{q0}$ modes   can be accomplished (Fig.~\ref{fig:SPADE})  in  the same way as  SPADE \cite{TNL16}. Mathematically, fin-SPADE implements a simultaneous measurement of the operators $\{\hat{N}_q = \hat{a}_q^\dag\, \hat{a}_q\}_{q=0}^Q$ with
\begin{align}
\hat{a}_q = \int_0^T \diff t \int_{\cl I} \diff\brho\,&\hat{E}(\brho,t)\,\psi_{q0}^*(\brho)\,\xi^*(t),
\end{align} 
resulting in  a $(Q+1)$-vector $\bm{N} = (N_0,\ldots, N_Q)^{\trans}$ of the number of counts in each mode. 

\begin{figure}[t] 
\centering\includegraphics[trim=18mm 66mm 25mm 75mm,width=0.7\columnwidth,clip=true]{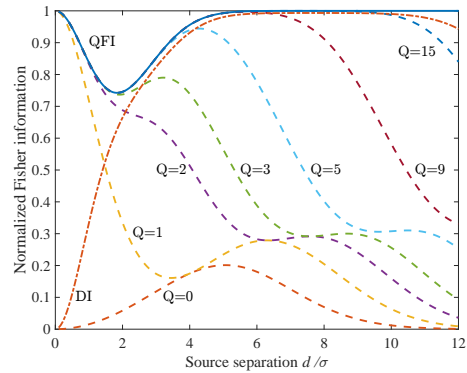}
\caption{\label{fig:Fin-SPADE} {Fin-SPADE performance:} The QFI (solid), the lower bound \eqref{eq:FIlb} on the FI of fin-SPADE (dashed) for various $Q$,  and of direct imaging (DI) (dashed-dotted). The Gaussian PSF \eqref{eq:gaussian} is assumed and $N_s =1.5$ photons. The plots are normalized to the maximum value $N_s/2\sigma^2$ of the QFI  and are independent of $\sigma$. The DI bound assumes a detector of width $17\sigma$ with $P_d=50$ pixels at 100\% fill factor and is stable to increase in $P_d$. Number-resolving unity-quantum-efficiency detectors are assumed for all the measurement schemes.} 
\end{figure}

The statistical correlations among the  HG modes in the state \eqref{eq:state}  make a direct calculation of the FI $\cl{J}_d[\bm{N}]$ of  fin-SPADE difficult. We turn instead to a  general lower bound on the  FI $\cl{J}_x[\bm{Y}]$ on  an arbitrary parameter $x$ of  any vector observation $\bm{Y} = (Y_1, \ldots, Y_M)^{\trans} \in \mathbb{R}^M$ depending on $x$.  For $\bm{\mu} = (\mean{Y_1}_x,\ldots, \mean{Y_M}_x)^{\trans}$ the mean vector and $\bm{C} = \mean{(\bm{Y} - \bm{\mu})(\bm{Y} - \bm{\mu})^{\trans}}_x $  the covariance matrix of $\bm{Y}$ evaluated at $x$, we have \cite{SMN14}:
\begin{align} \label{eq:FIlb}
\cl{J}_x[\bm{Y}] \geq \dot{\bm{\mu}}^{\textsf T}\, {\bm{C}}^{-1}\, \dot{\bm{\mu}},
\end{align}
where $\dot{\bm{\mu}}= \partial \bm{\mu}/\partial x$. Formally similar expressions have appeared in the quantum estimation literature \cite{HO04,*ZLJ+14}.

The mean and covariance  of $\bm{N}$ in the state $\rho_d$ for the fin-SPADE measurement can be calculated using semiclassical photodetection theory \cite{Sha09} as detailed in the Appendix. The resulting bound \eqref{eq:FIlb} is plotted in Fig.~\ref{fig:Fin-SPADE} for a representative value of $N_s = 1.5$ photons. Also shown is the lower bound \eqref{eq:FIlb} on the FI of spatially-resolved direct imaging (see also Figs.~\ref{fig:QFI} and \ref{fig:Pix-SLIVER} and the Appendix for details). 
Direct imaging is near quantum-optimal for $d \gtrsim 2\sigma$ -- in this regime, interference between the sources is minimal and the QCRB follows that for localizing a single source \cite{Hel76,TNL16}. We see that measuring the first 6 HG modes already achieves the quantum bound \eqref{eq:QFI} over the range $d = 0-4\sigma$ and that increasing $Q$ widens the region of saturation of the quantum bound.

\begin{figure}[t]
\centering\includegraphics[trim=8mm 27mm 14mm 10mm, width=0.8\columnwidth,clip=true]{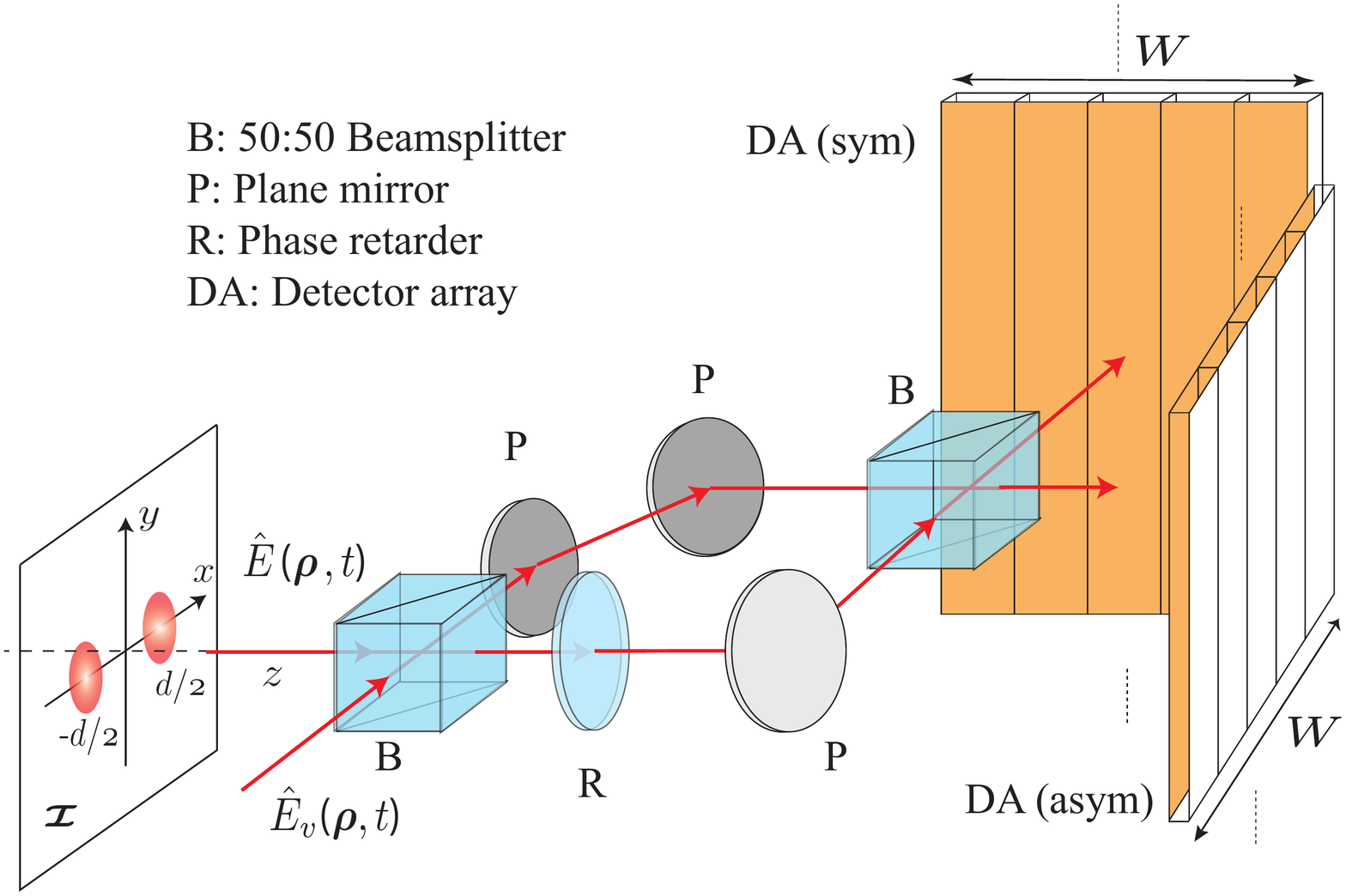}
\caption{\label{fig:SLIVER} Pix-SLIVER: The image-plane field is separated into its symmetric and antisymmetric components \eqref{eq:symasym} using a balanced Mach-Zehnder interferometer with an extra reflection in one arm before detecting the two outputs using identical detector arrays of width $W$ pixelated in the $x$-direction.} 
\end{figure}

\paragraph*{Pix-SLIVER:} Consider a PSF that is reflection-symmetric about the $y-$axis, i.e., $\psi(-x,y) = \psi(x,y)$, but otherwise arbitrary. Fig.~\ref{fig:SLIVER} shows a schematic of pix-SLIVER. Using an extra reflection in one arm of a balanced Mach-Zehnder interferometer, we separate the image-plane field into its  symmetric (\emph{s}) and antisymmetric (\emph{a}) components with respect to inversion of the image-plane field operator in the $x$-axis. The output field operators are 
\begin{align} \label{eq:symasym}
\hat{E}^{(s(a))}(x,y,t) = &\lb \hat{E}(x,y,t) \pm \hat{E}(-x,y,t)\rb/2  \nonumber\\
+&  \lb \hat{E}_v(x,y,t) \mp \hat{E}_v(-x,y,t)\rb/2,
\end{align}
where $\hat{E}_v(\brho,t)$ is the (vacuum-state) field operator entering the empty port of the first beam splitter in Fig.~\ref{fig:SLIVER}. The two outputs are detected using  two detector arrays pixelated along the $x$-direction. Each array consists of $P$ pixels of equal $x$-width. To show that super-resolution is  possible without number-resolving detectors, we assume \emph{on-off} detection in each pixel. For a pixel $p \in \{1,\ldots,P\}$  in the $\alpha\in \{s,a\}$ array, such a measurement corresponds to measuring the operator $\hat{K}^{(\alpha)}_p = f\lp \hat{N}_p^{(\alpha)}\rp$, where
\begin{align}
\hat{N}_p^{(\alpha)} = \int_0^T \diff t \int_{\cl{A}_p^{(\alpha)}} \diff \brho\, \hat{E}^{(\alpha)\dag}(\brho,t)\, \hat{E}^{(\alpha)}(\brho,t)
\end{align}
is the total photon number operator measured over the pixel area $\cl{A}_p^{(\alpha)}$ of array $\alpha$, and $f(x) = 0$ if $x=0$ and $1$ otherwise. The mean and covariance of the observation $\bm{K} = (K_1^{(s)},\ldots, K_P^{(s)}, K_1^{(a)}, \ldots,  K_P^{(a)})$ are calculated in the Appendix. For the Gaussian PSF \eqref{eq:gaussian}, the lower bound on the FI $\cl{J}_d[\bm{K}]$ of pix-SLIVER is plotted in Fig.~\ref{fig:Pix-SLIVER} for various values of $P$, showing how the QFI can be  approached more and more closely over the entire range of separation values by increasing $P$.

\begin{figure}[t]
\includegraphics[trim=0mm 68mm 6mm 74mm,width=0.9\columnwidth,clip=true]{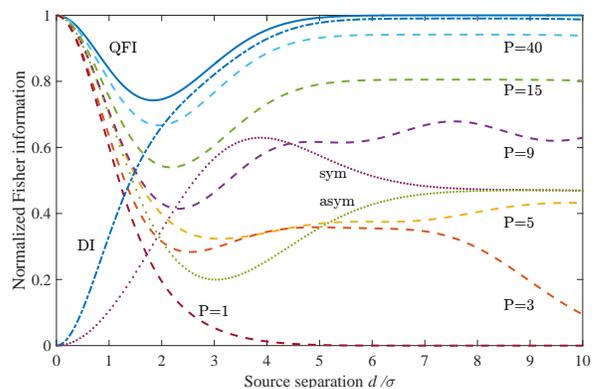}
\caption{\label{fig:Pix-SLIVER} {Pix-SLIVER performance:} The QFI (solid), the lower bound \eqref{eq:FIlb} on the FI of pix-SLIVER using on-off detection with various $P$ values(dashed lines),  the lower bound \eqref{eq:FIlb} for DI  (dash-dotted line) with number-resolved detection, and the contributions of the symmetric (sym) and antisymmetric (asym) field components to  \eqref{eq:FIlb}  for $P=40$ (dotted lines). The Gaussian PSF \eqref{eq:gaussian} is assumed and $N_s =1.5$ photons. The plots are normalized to the maximum value $N_s/2\sigma^2$ of the QFI  and are independent of $\sigma$. The lower bounds assume detector array(s) of width $17\sigma$ and 100\% fill factor. The DI bound assumes an array with $P_d=50$ pixels and is stable to increase in $P_d$. 
}
\end{figure}

\paragraph*{Discussion:}
The sensitivity of our schemes at sub-Rayleigh separations can be intuitively understood as follows. Information on $d$ is encoded in the energy distribution in any basis of spatial modes on $\cl{I}$, each of which is in a thermal state. The FI of any one mode scales roughly as $\sim \lb\overline{N}\,'(d)/\overline{N}(d)\rb^{2}$ \cite{NT16}, for $\overline{N}(d)$  the mean energy in the mode and $\overline{N}\,'(d) = \partial \overline{N}(d)/\partial d$, and is large if $\overline{N}(d) \sim 0$.  For fin-SPADE, while most of the energy is concentrated in the $\rm{TEM}_{00}$ mode, most of the FI is contributed by the $\rm{TEM}_{10}$ mode (Fig.~\ref{fig:Fin-SPADE}). Direct imaging is a poor way to estimate the energy in the latter, since the much larger energy in the $\rm{TEM}_{00}$ mode acts like background noise. Similarly, in pix-SLIVER, the antisymmetric component (comprising the odd modes in any basis of modes with definite parity about the centroid) carries the most information at sub-Rayleigh separations (Fig.~\ref{fig:Pix-SLIVER}).

 While the QCRB can be approached by the maximum-likelihood (ML) estimator in the limit of a large number of measurements \cite{VanTreesI},  suboptimal estimators  can also evade Rayleigh's curse \cite{TDL16,YTM+16,TFS16,PSH+16}.   That small values of $P$  achieve a substantial fraction of the QFI in pix-SLIVER is in line with work on detecting beam displacements using pixelated detectors \cite{KG14}.  The optical components used in pix-SLIVER have counterparts in other regions of the electromagnetic spectrum, leading to potential applications from the microwave to the gamma-ray regions \cite{[{}][{ (Planck Collaboration); }]PlanckColl15,*PS04,*Wei15,*Wee03}.  Generalizations to 2D-separation estimation \cite{ANT16arxiv} and variants of pix-SLIVER using image inversion devices \cite{WH07,*WSH09}, can be envisaged.  Recently developed techniques \cite{MM12,*BBP15} may help to generalize our quantum limit to multiple parameters and to unequal source strengths.

\begin{acknowledgements}
\paragraph*{Acknowledgements:} This work is supported by the Singapore National Research Foundation under NRF Grant No.~NRF-NRFF2011-07 and the Singapore Ministry of Education Academic Research Fund Tier 1 Project R-263-000-C06-112.

\paragraph*{Author contributions:} R.N. developed the source model, calculated the QFI, and invented pix-SLIVER. M.T. and R.N.  bounded the FI of fin-SPADE, and R.N. applied \eqref{eq:FIlb} to all the detection schemes.
\end{acknowledgements}

\paragraph*{Note added:} During this work, we became aware of an alternative derivation by Lupo and Pirandola \cite*{[{}][{ (to appear in Phys. Rev. Lett.)}]LP16arxiv} of a more general quantum bound applicable to arbitrary quantum states, including our bound Eq.~\eqref{eq:QFI} for thermal sources as a special case. Our proposal of concrete measurement schemes and their near-optimality for a broad range of source separations, however, are unique results here.

 %


\appendix
\section{Fundamental quantum limit on transverse resolution}

In this Section, we give the details of the derivation of the QCRB for separation estimation.  

As in Eq.~(4) of the main text, the quantum state of the electromagnetic field in the image plane is given by the coherent-state decomposition
\begin{align}\label{app:state}
\rho_d =  \int_{\mathbb{C}^2} \diff^2 A_+ \diff^2 A_-  \; P_{N_s}(A) \ket{\psi_{A,d}}\, \bra{\psi_{A,d}}.
\end{align}
Here
\begin{align} \label{eq:probamp}
P_{N_s}(A) =  \lp\frac{1}{\pi N_s}\rp^2 \, \exp\lp- \frac{\abs{A_+}^2 + \abs{A_-}^2}{N_s}\rp,
\end{align}
is the probability density of the source field amplitudes $A = (A_+,A_-)$ and the conditional state $ \ket{\psi_{A,d}}$ is an eigenvector of the image-plane field operator  $\hat{E}(\brho,t)$ with eigenfunction 
\begin{align}
\psi_{A,d}(\brho,t)= \lb A_+\,\psi(\brho - \bd/2) + {A}_- \, \psi(\brho + \bd/2) \rb\, \xi(t), \label{app:eigenfunc}
\end{align}
where $\bd = (d,0)$.
This eigenfunction is simply the semiclassical complex field amplitude  that results from the superposition of the images of the two sources conditioned on the amplitude vector $A$.

In order to evaluate the fidelity $F(\rho_{d_1}, \rho_{d_2})$ in Eq.~(6) of the main text, we need to first choose transverse spatial modes in which to express the quantum states $\rho_{d_1}$ and $\rho_{d_2}$.

\subsection{Transverse spatial modes}
For an arbitrary vector $\mb{a} = (a_x,a_y)$ in the image plane, consider the overlap function
\begin{align}
\delta(\mb{a}) :=  \int_{\cl I} \diff\brho\;\psi^*(\brho)\,\psi(\brho-\mb{a}).
\end{align}
The Cauchy-Schwarz inequality implies that $\abs{\delta(\mb{a})} \leq \delta(\mb{0}) = 1$.
We have
\begin{align}
\delta^*(\mb{a})
&=  \int_{\cl I} \diff\brho\;\psi^*(\brho-\mb{a})\,\psi(\brho) \\
&=  \int_{\cl I} \diff\brho\;\psi^*(\brho)\,\psi(\brho+\mb{a}) \\
&= \delta(-\mb{a}).
\end{align}
For an inversion-symmetric PSF, we can say more. Changing variables to $\bsigma = -\brho$ with $\diff\bsigma = \diff\brho$, we have
\begin{align}
\delta^*(\mb{a}) &= \int_{\cl I} \diff\bsigma\;\psi^*(-\bsigma)\,\psi(-\bsigma+\mb{a})\, \\
&= \int_{\cl I} \diff\bsigma\;\psi^*(\bsigma)\,\psi(\bsigma-\mb{a})\, \\
&\equiv \delta(\mb{a}),
\end{align}
where we have used  inversion-symmetry $\psi(-\brho) = \psi(\brho)$ of the PSF in the last step. For such PSFs, the overlap function is thus real-valued  for all $\mb{a} \in \cl{I}$. We make the inversion-symmetry assumption throughout this paper.

Since we are considering only the estimation of the $x$-component of the separation between the sources, we slightly abuse the above notation to define the overlap for a scalar argument as
\begin{align}
\delta(d) := \delta((d,0)).
\end{align}
We then have 
\begin{align} \label{app:deltaprops}
\delta(d) = \delta^*(d) = \delta(-d) \leq 1
\end{align}
for all values $d$ of the $x$-separation. 

 Consider two different values $d_1$ and $d_2$ of the separation. For $\bd_1 = (d_1,0)$, the functions
\begin{align} \label{chi12}
\begin{split}
\chi_1(\brho) &= \frac{\psi(\brho - \bd_1/2) + \psi(\brho +\bd_1/2)}{\sqrt{2\cl{N}_1}}\\
\chi_3(\brho) &= \frac{\psi(\brho - \bd_1/2) - \psi(\brho +\bd_1/2)}{\sqrt{2\,\cl{N}_3}}
\end{split}
\end{align}
 with normalization constants given by
\begin{align} \label{N13}
\cl{N}_1 & = {1 + \delta(d_1)}, \\
\cl{N}_3 & = {1 - \delta(d_1)},
\end{align}
are orthonormal over the image plane $\cl{I}$. The functions \eqref{chi12} will be two of our mode functions. 
In like manner, for $\bd_2 = (d_2,0)$, the functions
\begin{align}
\widetilde{\chi}_2(\brho) &= \frac{\psi(\brho - \bd_2/2) + \psi(\brho +\bd_2/2)}{\sqrt{2\,\cl{N}_2} }\\
\widetilde{\chi}_4(\brho) &= \frac{\psi(\brho - \bd_2/2) - \psi(\brho +\bd_2/2)}{\sqrt{2\,\cl{N}_4} }
\end{align}
are orthonormal over the image plane with the normalization constants 
\begin{align}
\cl{N}_2 & = {1 + \delta(d_2)}, \\
\cl{N}_4 & = {1 - \delta(d_2)}.
\end{align}
Using \eqref{app:deltaprops}, we can readily verify that $\widetilde{\chi}_2$ is  orthogonal to $\chi_3$ and $\widetilde{\chi}_4$ is  orthogonal to $\chi_1$. However $\widetilde{\chi}_2$ is not in general orthogonal to $\chi_1$ and neither is $\widetilde{\chi}_4$  orthogonal to $\chi_3$. In order to obtain an orthonormal set of transverse spatial modes,  the Gram-Schmidt process can be used to define
\begin{align}
{\chi}_2(\brho) &= \frac{\widetilde{\chi}_2(\brho) - \mu_s\,\chi_1(\brho) }{\sqrt{1-\mu^2_s}},\\
{\chi}_4(\brho) &= \frac{\widetilde{\chi}_4(\brho) - {\mu_a}\,\chi_3(\brho) }{\sqrt{1-\mu^2_a}},
\end{align}
with
\begin{align}
\mu_s &= \int_{\cl I} \diff\brho\, \chi_1^*(\brho) \widetilde{\chi}_2(\brho) = \frac{\delta\lb(d_1 - d_2)/2\rb + \delta\lb(d_1 + d_2)/2\rb}{\sqrt{\cl{N}_1\cl{N}_2}}, \\
\mu_a &= \int_{\cl I} \diff\brho\, \chi_3^*(\brho) \widetilde{\chi}_4(\brho) = \frac{\delta\lb(d_1 - d_2)/2\rb - \delta\lb(d_1 + d_2)/2\rb}{\sqrt{\cl{N}_3\cl{N}_4}}.
\end{align}
The set $\{\chi_1, \chi_2, \chi_3, \chi_4\}$ is  an orthonormal set of transverse spatial modes that span the same space as $\{\psi(\brho \pm \bd_1/2), \psi(\brho \pm \bd_2/2)\}$. Note that inversion symmetry of the PSF implies that $\chi_1$ and $\chi_2$ are symmetric with respect to inversion about $\brho=0$ while $\chi_3$ and $\chi_4$ are antisymmetric under inversion.

\subsection{Density operators $\rho_{d_1}$ and $\rho_{d_2}$} \label{app:densityops}
\noindent Equation \eqref{eq:probamp} implies that the incoherent thermal source amplitudes $A$ are circular-complex Gaussian random variables  satisfying the relations:
\begin{align} 
&\mathbb{E}[A_{\mu}] =  0 \label{S1}\\
&\mathbb{E}[A_{\mu}A_{\nu}] = 0 \label{S2}\\
&\mathbb{E}[A_{\mu}^*A_{\mu}] = N_s \label{S3} \\
&\mathbb{E}[A_{+}^*A_-] = 0 \label{S4}
\end{align}
for $\mu,\nu \in \{+,-\}$ ranging over the two sources.  Define the sum and difference amplitudes
\begin{align}
S &= A_+ + A_-, \\
D &= A_+ - A_-
\end{align}
which satisfy the relations
\begin{align} \label{SD}
\begin{split}
&\mathbb{E}[S] = \mathbb{E}[D] = 0 \\
&\mathbb{E}[S^2] =\mathbb{E}[D^2] = \mathbb{E}[SD] = 0 \\
&\mathbb{E}[S^{*} S] = \mathbb{E}[D^{*} D] = 2N_s \\
&\mathbb{E}[S^*D] = 0
\end{split}
\end{align}
and are thus \emph{statistically independent} circular-complex Gaussian random variables. Clearly, specifying the pair $(S,D)$ is equivalent to specifying $A=(A_+, A_-)$. The random variables $\abs{A_+}^2$ and $\abs{A_-}^2$ are independent and are both distributed exponentially with mean $N_s$ \cite{Goo85Statistical}. Analogously, the random variables $\abs{S}^2$ and $\abs{D}^2$ are also independent and are both distributed exponentially with mean $2N_s$.

Consider the coherent-state decomposition \eqref{app:state} for $\rho_{d_1}$. Conditioned on the source amplitudes, the eigenfunction \eqref{app:eigenfunc} can be rewritten as
\begin{align}
\psi_{A,d_1}(\brho,t) = \lp S\sqrt{\frac{\cl{N}_1 }{2}} \,\chi_1(\brho) + D\,\sqrt{\frac{\cl{N}_3 }{2}} \,\chi_3(\brho) \,\rp \xi(t),
\end{align}
in terms of the spatial modes defined in the previous subsection. Since $S$ and $D$ are i.i.d. circular-Gaussian variables, we may write, given the \emph{P}-representation \eqref{app:state} \cite{MW95,Sha09}:-
\begin{align}  \label{eq:rho1}
\rho_{d_1} = \rho_{\tsf{th}}\lp  \cl{N}_1N_s\rp \otimes \ket{0}\bra{0} \otimes \rho_{\tsf{th}}\lp \cl{N}_3N_s\rp\otimes \ket{0}\bra{0},
\end{align}
where
\begin{align}
\rho_{\tsf{th}}(\overline{N}) &= \sum_{n=0}^\infty \frac{{\overline{N}}^n}{(\overline{N}+1)^{n+1}} \ket{n}\bra{n} \\
& = \frac{1}{\pi \overline{N}} \int_{\mathbb{C}} \diff^2\alpha\, \exp\lp -\frac{\abs{\alpha}^2}{\overline{N}}\rp\,\ket{\alpha}\bra{\alpha} 
\end{align}
is the single-mode thermal state of $\overline{N}$ average photons (written above in its number-state and coherent-state decompositions). The four spatiotemporal modes in the above representation are respectively $\chi_1(\rho)\,\xi(t), \chi_2(\rho)\,\xi(t), \chi_3(\rho)\,\xi(t)$, and $\chi_4(\rho)\,\xi(t)$, and we have omitted including an infinity of other spatiotemporal modes which are in the vacuum state for all values of the separation and are not useful for estimating it.

Consider now the coherent-state decomposition \eqref{app:state} for $\rho_{d_2}$. Conditioned on the source amplitudes, the eigenfunction \eqref{app:eigenfunc} can be rewritten as
\begin{widetext}
\begin{align} \label{eq:rho2}
\psi_{A,d_2}(\brho,t) &= \lp S\sqrt{\frac{\cl{N}_2 }{2}} \,\widetilde{\chi}_2(\brho) + D\sqrt{\frac{\cl{N}_4 }{2}} \,\widetilde{\chi}_4(\brho)\rp \xi(t), \\
&=\left\{ S\sqrt{\frac{\cl{N}_2 }{2}} \,\lb\mu_s{\chi}_1(\brho) + \sqrt{1 -\mu_s^2} \,\chi_2(\brho)\rb+ D\sqrt{\frac{\cl{N}_4 }{2}} \,\lb\mu_a{\chi}_3(\brho) + \sqrt{1 -\mu_a^2}\, \chi_4(\brho) \rb \right\}\, \xi(t)
\end{align}
\end{widetext}
The unconditional density operator $\rho_{d_2}$ can then be written in the same set of modes used for writing \eqref{eq:rho1}, as follows:-
\begin{widetext}
\begin{align} \label{eq:rho2}
\rho_{d_2} =  \left\{{\Large U }_{s} \lb \rho_{\tsf{th}}\lp \cl{N}_2 N_s\rp  \otimes \ket{0}\bra{0} \rb {\Large U}_{s}^{\dag}\right\} \otimes \left\{ U_{a} \lb\rho_{\tsf{th}}\lp \cl{N}_4 N_s\rp \otimes \ket{0}\bra{0}\rb U_{a}^{\dag}\right\},
\end{align}
\end{widetext}
where $U_s$ is the two-mode beam-splitter unitary (see, e.g., ref.~\cite{GK05Quantum}) whose action on coherent states is
\begin{align}
U_s \Big( \ket{\alpha}\ket{\beta} \Big) \mapsto \ket{\mu_s\alpha - \sqrt{1 -\mu_s^2}\beta}\ket{\mu_s\beta + \sqrt{1 -\mu_s^2}\alpha} 
\end{align}
and on the number state-vacuum product is
\begin{align} \label{Usonn}
U_s \Big(\ket{n}\ket{0}\Big) \mapsto \sum_{k=0}^n \sqrt{\binom{n}{k}} \,\mu_s^k \,(1 - \mu_s^2)^{\frac{n-k}{2}} \ket{k}\ket{n-k}.
\end{align}
Similarly, $U_a$ is the two-mode beamsplitter unitary whose action on coherent states is
\begin{align}
U_a \Big(\ket{\alpha}\ket{\beta} \Big) \mapsto \ket{\mu_a\alpha + \sqrt{1 -\mu_a^2}\beta}\ket{\mu_a\beta - \sqrt{1 -\mu_a^2}\alpha} 
\end{align}
and on the number state-vacuum product is
\begin{align} \label{Uaonn}
U_a \Big( \ket{n}\ket{0}\Big) \mapsto \sum_{k=0}^n \sqrt{\binom{n}{k}} \,\mu_a^k \,(1 - \mu_a^2)^{\frac{n-k}{2}} \ket{k}\ket{n-k}.
\end{align}

\subsection{State fidelity}
\noindent The quantum fidelity between $\rho_{d_1}$ and $\rho_{d_2}$ is given by
\begin{align}
F(\rho_{d_1},\rho_{d_2}) = \Tr\sqrt{ \sqrt{\rho_{d_1}} \rho_{d_2} \sqrt{\rho_{d_1}} }. 
\end{align}
Since both density operators \eqref{eq:rho1} and \eqref{eq:rho2} factorize into a product of density operators on the symmetric (spanned by the modes $\chi_1(\rho)\,\xi(t)$ and $\chi_2(\rho)\,\xi(t)$) and the  antisymmetric modes (spanned by the modes $\chi_3(\rho)\,\xi(t)$ and $\chi_4(\rho)\,\xi(t)$), we can mutliply the fidelities for each pair. 

Considering the symmetric modes first, let
\begin{align}
r_1 &:= \frac{\cl{N}_1 N_s}{1 + \cl{N}_1 N_s}, \\
r_2 &:= \frac{\cl{N}_2 N_s}{1 + \cl{N}_2 N_s},
\end{align}
so that the symmetric components of the density operators under each hypothesis are
\begin{align}
\rho_{d_1}^{\mr{(sym)}} = (1-r_1)\sum_{n=0}^\infty r_1^n \ket{n}\bra{n}\otimes\ket{0}\bra{0}, \\
\rho_{d_2}^{\mr{(sym)}} = (1-r_2)\sum_{n=0}^\infty r_2^n\, U_s \Big( \ket{n}\bra{n}\otimes\ket{0}\bra{0} \Big) U_s^\dag.
\end{align}
Then
\begin{widetext}
\begin{align}
&\sqrt{\rho_{d_1}^{\mr{(sym)}}} \rho_{d_2}^{\mr{(sym)}} \sqrt{\rho_{d_1}^{\mr{(sym)}}} \\
&= (1-r_1)(1-r_2)\sum_{n,n',n''=0}^\infty r_1^{\frac{n+n''}{2}}\,r_2^{n'} \ket{n\,0}\bra{n\,0}\,U_s\,\ket{n'\,0}\bra{n'\,0}\,U_s^{\dag}\,\ket{n''\,0}\bra{n''\,0}, \label{Fsb}\\
&=(1-r_1)(1-r_2)\sum_{n,n',n''=0}^\infty r_1^{\frac{n+n''}{2}}\,r_2^{n'} \ket{n\,0}\mu_s^{n'}\,\delta_{n\,n'}\, \mu^{*n'}_s \, \delta_{n\,n''}\bra{n''\,0} \\
&=(1-r_1)(1-r_2)\sum_{n=0}^\infty r_1^{n}\,r_2^{n} \abs{\mu_s}^{2n} \ket{n\,0}\bra{n\,0},
\end{align}
where we have used Eq.~\eqref{Usonn} to evaluate the matrix elements in Eq.~\eqref{Fsb}. Consequently,
\begin{align} \label{Fsym}
F \lp\rho_{d_1}^{\mr{(sym)}},\rho_{d_2}^{\mr{(sym)}} \rp&= \Tr \sqrt{ \sqrt{\rho_{d_1}^{\mr{(sym)}}} \rho_{d_2}^{\mr{(sym)}} \sqrt{\rho_{d_1}^{\mr{(sym)}}}} \\
&= \frac{(1-r_1)^{1/2}(1-r_2)^{1/2}}{1 - \abs{\mu_s}\sqrt{r_1\,r_2}}\\
&= \lb \sqrt{\lp 1 + \cl{N}_1N_s\rp \lp 1 + \cl{N}_2 N_s\rp} - \abs{\mu_s}\sqrt{\cl{N}_1\,\cl{N}_2}\,N_s\rb^{-1}.
\end{align}
In similar fashion, we find
\begin{align} \label{Fasym}
F \lp\rho_{d_1}^{\mr{(asym)}},\rho_{d_2}^{\mr{(asym)}} \rp&=\lb \sqrt{\lp 1 + \cl{N}_3N_s\rp \lp 1 + \cl{N}_4 N_s\rp} - \abs{\mu_a}\sqrt{\cl{N}_3\,\cl{N}_4}\,N_s\rb^{-1},
\end{align}
resulting in  the expression
\begin{align}
F(\rho_{d_1},\rho_{d_2}) &=
\lb \sqrt{\lp 1+ N_s\,[1+\delta(d_1)]\rp \lp 1+ N_s[1+\delta(d_2)] \rp} - N_s \Big\vert\delta\lb(d_1 - d_2)/2\rb + \delta\lb(d_1 + d_2)/2\rb \Big\vert \rb^{-1} \nonumber \\
&\times
\lb \sqrt{\lp 1+ N_s\,[1-\delta(d_1)]\rp \lp 1+ N_s\,[1-\delta(d_2)]\rp} - N_s \Big\vert\delta\lb(d_1 - d_2)/2\rb - \delta\lb(d_1 + d_2)/2\rb\Big\vert \rb^{-1}
\end{align}
for the overall fidelity.
\end{widetext}

\subsection{Quantum Cram\'er-Rao bound}
\noindent Let $d_1 = d$ and $d_2 =d_1 + \Delta d$. The quantum Fisher information (QFI) $\cl{K}_d$ on $d$ is given by \cite{BC94,Hay06}
\begin{align}
\cl{K}_d = 8 \times \lim_{\Delta d \rightarrow 0}\frac{1 - F(\rho_{d},\rho_{d+\Delta d})} {\lp\Delta d \rp^2} = -\hspace{1mm}4 \frac{\partial^2 F(\rho_{d_1},\rho_{d_2}) }{\partial d_2^2}\bigg\vert_{d_2 = d_1}.
\end{align}
Since the symmetric and antisymmetric modes are in  tensor-product states, $\cl{K}_d$ is the sum of the QFIs $\cl{K}_d^{\mr{sym}}$ and $\cl{K}_d^{\mr{asym}}$ from the respective subsystems \cite{Hay06}.  Defining 
\begin{align}
\begin{split}
\gamma(d) &= \delta'(d), \\
\beta(d) &=\gamma'(d),
\end{split}
\end{align}
 the QFI from the symmetric modes is found after some algebra  to be:
\begin{align} \label{Ksym}
\cl{K}_d^{\mr{sym}} = [ \beta(d) - \beta(0)]N_s - \frac{N_s^2\,\gamma^2(d)}{1 + N_s[1+\delta(d)]}.
\end{align}
Similarly, the QFI from the antisymmetric modes is found to be
\begin{align} \label{Ksym}
\cl{K}_d^{\mr{asym}} = -[ \beta(d) + \beta(0)]N_s - \frac{N_s^2\,\gamma^2(d)}{1 + N_s[1-\delta(d)]},
\end{align}
giving a total QFI
\begin{align} \label{QFIsupp}
\cl{K}_d &= \cl{K}_d^{\mr{sym}} + \cl{K}_d^{\mr{asym}} \\
&= - 2\beta(0)N_s - 2\gamma^2(d) \lb\frac{(1+ N_s)N_s^2}{(1+N_s)^2 - N_s^2\delta^2(d)} \rb,
\end{align}
which is Eq.~(7) of the main text. Here
\begin{align}
\beta(0) = - \int_{\cl I} \diff\brho \abs{\frac{\partial \psi(\brho)}{\partial x}}^2 \equiv -(\Delta k_x^2),
\end{align}
For circularly symmetric PSFs, this quantity is independent of the direction of the $x$-axis and is the mean-squared spatial bandwidth of the PSF.  

\section{Fisher Information lower bounds for concrete measurements}

In this Section, we give the derivation of the lower bound on the Fisher information for  direct imaging, fin-SPADE, and pix-SLIVER.

Consider a vector random variable $\bm{Y} = (Y_1,\ldots,Y_M)^{\trans} \in \mathbb{R}^M$ whose probability density $P_{\bm{Y}|X}(\bm{y}|x)$ depends on an unknown parameter $x$. The classical Fisher information (FI) $\cl{J}_x[\bm{Y}]$ of $\bm{Y}$ on $x$ \cite{VanTreesI} is typically difficult to compute unless the components of $\bm{Y}$ are statistically independent. However, a  general \emph{lower bound}
\begin{align} \label{app:FIlb}
\cl{J}_x[\bm{Y}] \geq \dot{\bm{\mu}}^{\textsf T}\, {\bm{C}}^{-1}\, \dot{\bm{\mu}}
\end{align}
was recently derived in \cite{SMN14}. Here $\bm{\mu} = (\mean{Y_1}_x,\ldots, \mean{Y_M}_x)^{\trans}$ is the mean observation vector, $\bm{C} = \mean{(\bm{Y} - \bm{\mu})(\bm{Y} - \bm{\mu})^{\trans}}_x $  is the covariance matrix of $\bm{Y}$, and $\dot{\bm{\mu}}= \partial \bm{\mu}/\partial x$. All the above quantities are functions of $x$. The bound \eqref{app:FIlb} is very convenient as it depends only on the first two moments of the observation vector, which are easier to compute. In contrast, the FI $\cl{J}_x[\bm{Y}]$ depends on the full joint probability density of $Y$ (conditioned on $x$).

We compute this lower bound for various measurements below. Since all the measurements involve at most linear-optical processing prior to photodetection, the classicality (in the sense of having a non-negative $P$-representation \cite{MW95,Sha09}) of the incoming state $\rho_d$ is preserved. It is well known that, for such states, the quantum theory of photodetection gives the same quantitative statistics as the semiclassical theory of photodetection \cite{MW95,Sha09}.  Let  the input field $E(\brho,t)$   be subjected to arbitrary linear-optics processing and the resulting output field $E_{\tsf{det}}(\brho,t)$ impinge on an ideal continuum photodetector surface. Semiclassical photodetection theory dictates that, conditioned on the source amplitudes $A$, the incident field generates a space-time Poisson random process at the photodetector output with the rate function (or intensity)
$\abs{E_{\tsf{det}}(\brho,t)}^2$. Unconditional statistics can then be obtained by averaging over the source distribution using \eqref{eq:probamp}. We will follow this approach in the sequel.

\subsection{\label{sec:di} Lower bound on direct imaging}
Consider first the case of direct detection in the image plane with a pixelated detector array centered at the origin and of width $W$ in the $x$-direction. For simplicity, we assume it to be infinite in the $y$-direction, but pixelated in the $x$-direction with $P_d$ pixels of width $W/{P_d}$. We assume ideal unity-quantum-efficiency and noiseless number-resolved photon counting in each pixel. Let $p \in \{1,\ldots, P_d\}$ be the pixel index and let pixel $p$ be defined by the region
\begin{align} \label{eq:Ap}
\cl{A}_p = \{(x,y) : l_p \leq x \leq r_p, -\infty \leq y \leq\infty\}
\end{align}
of the image plane. The observation consists of the vector $\bm{N} = (N_1,\ldots, N
_{P_d})^{\trans}$ of measured counts in each pixel.

Conditioned on $A$, the intensity function $I_A(\brho,t)$ in the image plane is, using \eqref{app:eigenfunc},
\begin{widetext}
\begin{align}
&I_A(\brho,t) = \abs{\psi_{A,d}(\brho,t)}^2 \\
&= \left\{ \abs{A_+}^2\,\abs{\psi(\brho - \bd/2) }^2+ \abs{{A}_-}^2 \, \abs{\psi(\brho - \bd/2)}^2  + 2 {\rm{Re}} \lb {A_+^*\, A_-}\, \psi^*(\brho - \bd/2)\,\psi(\brho + \bd/2)\rb \right\}\abs{\xi(t)}^2. 
\end{align}
\end{widetext}
The conditional photocounts $N_{p|A}$ on the detectors $p \in \{1,\ldots,P_d\}$ integrated over the observation interval $[0,T]$ are then independent Poisson random variables with the means
\begin{align}
\mu_{p|A} = \int_0^T \diff t\int_{\cl{A}_p} \diff \brho\, I_A(\brho,t).
\end{align}
We now suppose the PSF has the Gaussian form 
\begin{align} \label{app:gaussian}
\psi_G(\brho) =\frac{1}{(2\pi\sigma^2)^{1/2}}\exp\lp-\frac{\abs{\brho}^2}{4\sigma^2}\rp,
\end{align}
although the treatment is readily generalized to arbitrary PSFs. We obtain
\begin{align} \label{eq:mupA}
\mu_{p|A} = \abs{A_+}^2 \,\alpha_p + 2 {\rm{Re}}\lp A_+^*\, A_-\rp \beta_p + \abs{A_-}^2 \gamma_p,
\end{align}
where
\begin{align}
\alpha_p &= Q\lp \frac{l_p +d/2}{\sigma}\rp - Q\lp \frac{r_p +d/2}{\sigma}\rp, \nonumber\\ 
\beta_p &= 2\exp \lp \frac{-d^2}{8\sigma^2}\rp \,\lb Q\lp \frac{l_p }{\sigma}\rp - Q\lp \frac{r_p }{\sigma}\rp\rb,  \label{eq:alphabetagamma}\\
\gamma_p &=Q\lp \frac{l_p  - d/2}{\sigma}\rp - Q\lp \frac{r_p  - d/2}{\sigma}\rp, \nonumber
\end{align}
and
\begin{align}
Q(x) = \frac{1}{\sqrt{2\pi}}\int_{x}^{\infty} \diff t\,\exp \lp \frac{-t^2}{2}\rp
\end{align}
is the Q-function. 

The mean photocount $\mu_p = \mathbb{E}[N_p]= \mathbb{E}_A [\mu_{p|A}]$ is then
\begin{align} \label{eq:mudi}
\mu_p = N_s (\alpha_p + \gamma_p),
\end{align}
where we have used eqs.~\eqref{S1}-\eqref{S4}. We then have
\begin{widetext}
\begin{align} \label{eq:mudidot}
\dot{\mu}_p &= \frac{\partial \mu_p}{\partial d} \nonumber \\
&= \frac{N_s}{2\sqrt{2\pi}\sigma} \left\{ \exp\lb \frac{-\lp l_p -d/2\rp^2}{2\sigma^2}\rb - \exp\lb \frac{-\lp r_p -d/2\rp^2}{2\sigma^2}\rb +\exp\lb \frac{-\lp r_p +d/2\rp^2}{2\sigma^2}\rb - \exp\lb \frac{-\lp l_p +d/2\rp^2}{2\sigma^2}\rb\right\}.
\end{align}
\end{widetext}
The $(p,p')$-th element of the covariance matrix of $\bm{N}$ equals $\mathbb{E}[N_pN_{p'}] - \mu_p \,\mu_{p'}$. Now
\begin{align}
\mathbb{E}[N_pN_{p'}] &= \mathbb{E}_A[\mu_{p|A}\;\mu_{p'|A}] \\
&=\left\{
	\begin{array}{ll}
		\mathbb{E}_A[\mu_{p|A}] \mathbb{E}_A[\mu_{p'|A}] & \mbox{if } p \neq p'  \\
		\mathbb{E}_A[\mu^2_{p|A}] & \mbox{if } p=p'.
	\end{array}
\right.
\end{align}
Straightforward computations using the relations \eqref{S1}-\eqref{S4} and \eqref{eq:mupA} give the matrix elements
\begin{align} \label{eq:covdi}
 C_{p\,p'} = \left\{
	\begin{array}{ll}
		N_s^2(\alpha_p^2+ 2 \beta_p^2 + \gamma_p^2) + N_s(\alpha_p + \gamma_p) & \mbox{if } p =p', \\
		 N_s^2(\alpha_p\alpha_{p'} + 2 \beta_p\,\beta_{p'} + \gamma_p\,\gamma_{p'})  & \mbox{if } p \neq p'. 
 	\end{array}
\right.
\end{align}
In obtaiing the elements of the covariance matrix, we have also used the fact that $\mathbb{E}[\abs{A_+}^4] = \mathbb{E}[\abs{A_-}^4] = 2N_s^2$, which follows from the exponential statistics of $\abs{A_+}^2$ and $\abs{A_-}^2$ (see Sec.~\ref{app:densityops}). Using eqs.~\eqref{eq:mudidot} and \eqref{eq:covdi}, the lower bound \eqref{app:FIlb} can be evaluated numerically for any given system parameters -- see Figs.~2, 4, and 6 of the main text. The limit of continuum image-plane photodetection is achieved for $P_d \rightarrow \infty$, but it was observed that the the FI lower bound did not change discernibly for $P_d \gtrsim 50$, so $P_d=50$ was used in plotting the direct imaging curves in Figs.~2, 4, and 6 of the main text.

\subsection{Lower bound on fin-SPADE performance}
Suppose the PSF has the Gaussian form 
\begin{align} \label{app:gaussian}
\psi_G(\brho) =\frac{1}{(2\pi\sigma^2)^{1/2}}\exp\lp-\frac{\abs{\brho}^2}{4\sigma^2}\rp.
\end{align}
As discussed in the main text, the fin-SPADE measurement measures the photon number in each Hermite-Gaussian mode $\mr{TEM}_{q0}$ (with profile $\psi_{q0}(\brho)$) of the image-plane field for $0\leq q \leq Q$ over the interval $[0,T]$. This  results in a $(Q+1)$-vector  $\bm{N} = (N_0,\ldots, N_Q)^{\trans}$ of the number of counts in each mode. The moments of $\bm{N}$ can be found using the semiclassical photodetection theory as follows.

Conditioned on $A$, the amplitude $B_{q|A}$ in the $q$-th channel can be written (cf. Eq.~(11) of the main text):-
\begin{align}
B_{q|A} &= \int_0^T \diff t \int_{\cl I} \diff \brho \,\psi_{A,d}(\brho,t) \,\psi^*_{q0}(\brho)\,\xi^*(t). 
\end{align}

As shown in \cite{TNL16}, the integrals may be associated to the probability amplitudes of a coherent state in the Fock basis so that
\begin{align} \label{eq:amplitudes}
B_{q|A} = 
		\frac{\kappa^{q/2} \exp(-\kappa/2)}{\sqrt{q!}}\; R_q, 
\end{align}
where 
\begin{align} 
R_{q} = \left\{
	\begin{array}{ll}
		 S   & \mbox{(if $q$ even)}   \\
		 D   & \mbox{(if $q$ odd)},
	\end{array}
\right.
\end{align}
and
\begin{align} \label{kappa}
\kappa = \frac{d^2}{16\sigma^2} .
\end{align}

Conditioned on $A$, the photocounts $N_{q|A}$ in each $q$-channel are  independent Poisson random variables with the means
\begin{align}
\mu_{q|A} &= \abs{B_{q|A}}^2 =\frac{\kappa^{q} \exp(-\kappa)}{q!}\; \abs{R_q}^2 \\
&\equiv f_q \abs{R_q}^2,
\end{align}
where $f_q$ is the Poisson probability of mean $\kappa$. For the unconditional mean, we have
\begin{align}
\mu_q :=\mean{N_q} &= \mathbb{E}_A \lb {\mu_{q|A}} \rb \\
&=\mathbb{E}_A[  f_q \abs{R_q}^2] \\
& = 2N_s f_q, \label{mean}
\end{align}
since $\abs{S}^2$ and $\abs{D}^2$ are i.i.d. random variables distributed exponentially with mean $2N_s$.
 We also need
\begin{align} \label{eq:spademudot}
\frac{\partial \mu_q}{\partial d} &= \frac{N_s d}{4\sigma^2}\, \frac{\kappa^{q-1} [q-\kappa]\exp(-\kappa)}{q!} \\
&=  \frac{N_s d}{4\sigma^2} (f_{q-1}-f_q),
\end{align}
where we define $f_{-1} = 0$.

For the second moments, three cases arise. First, for $q = {q'}$, we have
\begin{align}
\mathbb{E}[{N_q^2}] &= \mathbb{E}_A\lb \mathbb{E}[N_{q|A}^2]\rb  \\
		&=\mathbb{E}_A\lb f_q^2 \abs{R_q}^4+ f_q \abs{R_q}^2\rb\\
& = 8N_s^2\,f_q^2 + 2N_s f_q,     
\end{align}
where we have used the fact that $N_{q|A}$ is Poisson-distributed.
If $q \neq {q'}$ but $q-{q'}$ is even, $R_q = R_{q'}$, so we get
\begin{align}
\mathbb{E}[N_q\,N_{q'}] &= \mathbb{E}_A\lb \mathbb{E}[N_{q|A}N_{q'|A}]\rb  \\
				  &= \mathbb{E}_A\lb \mu_{q|A}\, \mu_{q'|A}\rb  \\
				  &=\mathbb{E}_A\lb f_q \,f_{q'} \abs{R_q}^4 \rb\\
				& = 8N_s^2\,f_q\,f_{q'} . 
\end{align}
If $q \neq {q'}$ and $q-{q'}$ is odd, $\mathbb{E}_A[\abs{R_q}^2\,\abs{R_{q'}}^2]= \mathbb{E}_A[\abs{R_q}^2]\,\mathbb{E}_A[\abs{R_{q'}}^2]$, so that
\begin{align}
\mathbb{E}[N_q\,N_{q'}] &= \mathbb{E}_A\lb \mathbb{E}[N_{q|A}N_{q'|A}]\rb  \\
				  &= \mathbb{E}_A\lb \mu_{q|A}\, \mu_{q'|A}\rb  \\
				  &=\mathbb{E}_A\lb f_q\, f_{q'} \abs{R_q}^2 \abs{R_{q'}}^2 \rb\\
				& = 4N_s^2\,f_q\,f_{q'} . 
\end{align}
 Thus, the covariance matrix $\bm{C}$ of $\bm{N}$ has  the $q{q'}-\mbox{th}$ entry
\begin{align} \label{eq:spadecov}
C_{q{q'}} &= \left\{
	\begin{array}{ll}
		4N_s^2\,f_q^2 + 2N_s f_q  & \mbox{if } q={q'},  \\
		4N_s^2\,f_q\,f_{q'} & \mbox{if } q \neq {q'} \mbox{ and } {q-{q'}} \mbox{ is even,}   \\
           0&  \mbox{if } q \neq {q'} \mbox{ and } {q-{q'}} \mbox{ is odd.}
	\end{array}
\right.
\end{align}
From Eqs.~\eqref{eq:spademudot} and \eqref{eq:spadecov}, the lower bound \eqref{app:FIlb} can be numerically evaluated, as displayed in Fig.~4 of the main text.

\subsection{Lower bound on pix-SLIVER performance}
Consider the pix-SLIVER setup of Fig.~5 of the main text with identical detector arrays in the symmetric (s) and antisymmetric (a) output ports. The overall dimensions of the arrays are as in Sec.~\ref{sec:di}, except that we consider $P$ pixels in each array. For a conservative comparison, we take $P < P_d$. In addition, we also assume on-off (Geiger mode) detection in each pixel, so that each component of the observation $\bm{K} = (K_1^{(s)},\ldots, K_P^{(s)}, K_1^{(a)}, \ldots,  K_P^{(a)})$ is 0 (if the corresponding pixel did not fire) or 1 (if it did). In contrast, we allowed number-resolved detection in direct imaging (see Sec.~\ref{sec:di}).

We now assume that the PSF is symmetric relative to \emph{reflection} about the $y-$axis, i.e., $\psi(-x,y) = \psi(x,y)$ for all $x$ and $y$ -- circular symmetry of the PSF is clearly a sufficient condition for this to hold. Conditioned on $A$, the semiclassical field amplitude in the two interferometer outputs is given by (cf. Eq.~(13) of the main text):-
\begin{align}
E^{\lp s \lp a \rp \rp}_{A}(x,y,t) &= \lb \psi_{A,d}(x,y,t) \pm \psi_{A,d}(-x,y,t)\rb/2.
\end{align} 
Since the field $\hat{E}_v(\brho,t)$ is in vacuum, the open input port of the first beam splitter does not contribute to the field amplitude. We can rewrite the above as
\begin{align}
E^{(s)}_{A}(x,y,t) &=\frac{S}{2} \lb \psi(x+d/2,y,t) + \psi(x-d/2,y,t)\rb, \\
E^{(a)}_{A}(x,y,t) &=\frac{D}{2} \lb \psi(x-d/2,y,t) - \psi(x+d/2,y,t)\rb.
\end{align}
where we have used the reflection symmetry of the PSF. The resulting conditional intensity patterns on the two detectors are
\begin{align}
I^{(s)}_{A}(x,y,t) &=\frac{\abs{S}^2}{4} \lb \abs{\psi(x-d/2,y,t) }^2+  \abs{\psi(x+d/2,y,t)}^2 \rb \nonumber \\
& + \frac{\abs{S}^2}{2} \mr{Re}\lb \psi^*(x - d/2,y,t)\,\psi(x+d/2,y,t) \rb, \\
I^{(a)}_{A}(x,y,t) &=\frac{\abs{D}^2}{4} \lb \abs{\psi(x-d/2,y,t) }^2+  \abs{\psi(x+d/2,y,t)}^2  \rb \nonumber \\ 
&-\frac{\abs{D}^2}{2} \mr{Re}\lb \psi^*(x - d/2,y,t)\,\psi(x+d/2,y,t) \rb.
\end{align}
The integrated intensity $I^{(\alpha)}_{p|A}$ on  pixel $p \in \{1,\ldots,P\}$ of the $\alpha \in \{s,a\}$ detector array over the observation interval $[0,T]$ is then
\begin{align}
I^{(\alpha)}_{p|A} = \int_0^T \diff t \int_{\cl{A}_p} \diff \brho \,I^{(\alpha)}_{A}(x,y,t). 
\end{align}

Specializing to the Gaussian PSF \eqref{app:gaussian}, these integrals evaluate to
\begin{align}
I^{(s)}_{p|A} &= \frac{\abs{S}^2}{4} \left [\alpha_p + \gamma_p + \beta_p\right] \equiv \frac{\abs{S}^2}{4}f_p^{(s)}, \\
I^{(a)}_{p|A} &= \frac{\abs{D}^2}{4} \left[ \alpha_p + \gamma_p - \beta_p\right] \equiv \frac{\abs{D}^2}{4} f_p^{(a)},
\end{align}
where $\alpha_p$, $\gamma_p$, and $\beta_p$ are defined in Eq.~\eqref{eq:alphabetagamma} and the above equations serve to define the quantities $\{f_p^{(\alpha)}\}$.

Conditioned on $A$, the probability of a detector click in the $(\alpha,p)$-th pixel is simply the probability that one or more photons impinge on the pixel:
\begin{align}
\mathbb{E}[K_{p|A}^{\alpha}]  \equiv \mu_{p|A}^{(\alpha)}=1 - \exp(-I^{(\alpha)}_{p|A}).
\end{align}
 Consequently,
\begin{align}
\mu_{p}^{(\alpha)} &\equiv  \mathbb{E}[K_p^{\alpha}] \\
&= \mathbb{E}_A\lb K_{p|A}^{(\alpha)}\rb \\
&= 1 - \mathbb{E}_A\lb\exp(-I^{(\alpha)}_{p|A})\rb \\
&= \frac{f_p^{(\alpha)}N_s}{2+f_p^{(\alpha)}N_s}, \label{eq:slivermu}
\end{align}
where we have used the fact that $|S|^2$ and $|D|^2$ are exponentially distributed with mean $2N_s$ to evaluate the expectation over $A$. It follows that
\begin{align} \label{eq:slivermudot}
\dot{\mu}_{p}^{(\alpha)}= \frac{2\dot{f}_p^{(\alpha)}N_s}{\lp2+f_p^{(\alpha)}N_s\rp^2},
\end{align}
for

\begin{widetext}
\begin{align}
\dot{f}_{p}^{(s (a))} & = \frac{1}{2\sqrt{2\pi}\sigma} \left\{ \exp\lb \frac{-\lp l_p -d/2\rp^2}{2\sigma^2}\rb - \exp\lb \frac{-\lp r_p  -d/2\rp^2}{2\sigma^2}\rb +\exp\lb \frac{-\lp r_p +d/2\rp^2}{2\sigma^2}\rb - \exp\lb \frac{-\lp l_p +d/2\rp^2}{2\sigma^2}\rb \right\} \nonumber \\
 &\mp \lp \frac{d}{2\sqrt{2\pi}\sigma^2}\rp \exp \lp \frac{-d^2}{8\sigma^2}\rp \,\lb Q\lp \frac{l_p }{\sigma}\rp - Q\lp \frac{r_p }{\sigma}\rp\rb.
\end{align}

For the second moments $\mathbb{E}\lb K_p^{(\alpha)}\, K_{p'}^{(\alpha')}\rb$, three cases arise. If $p=p'$ and $\alpha = \alpha'$,
\begin{align}
\mathbb{E}\lb K_p^{(\alpha)}\, K_{p'}^{(\alpha')}\rb 
&=\mathbb{E}\lb K_p^{(\alpha)}\rb \\
&= \mathbb{E}_A[\mu_{p|A}^{(\alpha)} ] \\
&= \mu_{p}^{(\alpha)}.
\end{align}
If $\alpha \neq \alpha'$ (so that the pixels are in different detector arrays), the independence of $S$ and $D$ ensures that $ K_p^{(\alpha)}$ and $ K_{p'}^{(\alpha')}$ are independent also so that
\begin{align} \label{app:diffarrays}
\mathbb{E}\lb K_p^{(\alpha)}\, K_{p'}^{(\alpha')}\rb 
&= \mu_{p}^{(\alpha)} \mu_{p'}^{(\alpha')}.
\end{align}
Finally, if $\alpha = \alpha'$ but $p \neq p'$, 
\begin{align}
\mathbb{E}\lb K_p^{(\alpha)}\, K_{p'}^{(\alpha')}\rb 
&= \mathbb{E}_A \lb \mathbb{E} \lb K_{p|A}^{(\alpha)}\, K_{p'|A}^{(\alpha)} \rb\rb \\
&= \mathbb{E}_A \lb \mu_{p|A}^{(\alpha)}\, \mu_{p'|A}^{(\alpha)}\rb \\
&=  \mathbb{E}_A \lb  \lp 1 - \exp(-I^{(\alpha)}_{p|A})\rp \lp 1 - \exp(-I^{(\alpha)}_{p'|A})\rp \rb \\
&= 1 - \frac{2}{2+f_p^{(\alpha)}N_s} - \frac{2}{2+f_{p'}^{(\alpha)}N_s} + \frac{2}{2+\lp f_{p}^{(\alpha)} + f_{p'}^{(\alpha)} \rp N_s},
\end{align}
\end{widetext}
where we again use the exponential distribution of $|S|^2$ and $|D|^2$ to evaluate the expectation over $A$.
From these second moments, means \eqref{eq:slivermu}, and \eqref{eq:slivermudot}, the lower bound \eqref{app:FIlb} can be numerically evaluated, with the results displayed in Fig.~6 of the main text. Note that Eq.~\eqref{app:diffarrays} implies that the covariance matrix $\bm{C}$ is a direct sum of matrices for the symmetric and antisymmetric outputs, so that the lower bound \eqref{app:FIlb} is also the sum of corresponding terms -- these are shown separately in Fig.~6 of the main text for the case of $P=40$.


\end{document}